\documentclass{myPoS}
\usepackage{amsmath} 
\usepackage{amssymb}
\usepackage{mathrsfs}
\usepackage{graphicx}
\usepackage{graphics}

\newcommand{\be}{\begin{equation}}
\newcommand{\ee}{\end{equation}}
\newcommand{\ba}{\begin{eqnarray}}
\newcommand{\ea}{\end{eqnarray}}
\def\beann{\begin{eqnarray*}} \def\eeann{\end{eqnarray*}}
\newcommand{\bal}{\begin{align}}
\newcommand{\eal}{\end{align}}

\def\slash#1{#1\!\!\!\!\!/\!\,\,}
\def\Dslash{\slash D}

\def\lsim{\raise0.3ex\hbox{$<$\kern-0.75em\raise-1.1ex\hbox{$\sim$}}}
\def\gsim{\raise0.3ex\hbox{$>$\kern-0.75em\raise-1.1ex\hbox{$\sim$}}}

\title{
\vspace*{-3cm}
\begin{flushright}\texttt{\footnotesize
CERN-PH-TH/2008-221}
\end{flushright}
\vfill
The curvature of the critical surface $(m_{u,d},m_s)^{\rm crit}(\mu)$: \\
a progress report} 

\ShortTitle{Curvature of critical surface} 

\author{\speaker{Philippe de Forcrand} \\ 
Institute for Theoretical Physics, ETH Zurich, CH-8093 Zurich, Switzerland\\
and\\
CERN, Physics Department, TH Unit, CH-1211 Geneva 23, Switzerland\\ 
E-mail: \email{forcrand@phys.ethz.ch}} 

\author{Owe Philipsen\\
        Institut f\"ur Theoretische Physik, Westf\"alische Wilhelms-Universit\"a
t M\"unster, Germany \\
        E-mail: \email{ophil@uni-muenster.de}}

\abstract{
At zero chemical potential $\mu$, the order of the temperature-driven 
quark-hadron transition
depends on the quark masses $m_{u,d}$ and $m_s$. Along a critical line 
bounding the region of first-order chiral transitions in the
$(m_{u,d},m_s)$ plane, this transition is second order. When the chemical
potential is turned on, this critical line spans a surface, whose curvature
at $\mu=0$ can be determined without any sign or overlap problem.
Our past measurements on $N_t=4$ lattices suggest that the region of quark
masses for which the transition is first order {\em shrinks} when $\mu$ is
turned on, which makes a QCD chiral critical point at small $\mu/T$ unlikely.
We present results from two complementary methods, which can be combined to
yield information on higher-order terms. It turns out that the ${\cal O}(\mu^4)$
term reinforces the effect of the leading ${\cal O}(\mu^2)$ term,
and there is strong evidence that the ${\cal O}(\mu^6)$ and ${\cal O}(\mu^8)$ terms
do as well.
We also report on simulations underway, where the strange quark is given its 
physical mass, and where the lattice spacing is reduced.
} 

\FullConference{The XXVI International Symposium on Lattice Field Theory \\ July 14 - 19, 2008\\ Williamsburg, Virginia, USA} 

\begin{document} 

\section{Introduction}

\begin{figure}[t]
\begin{center}
\centerline{
\includegraphics[width=0.30\textwidth]{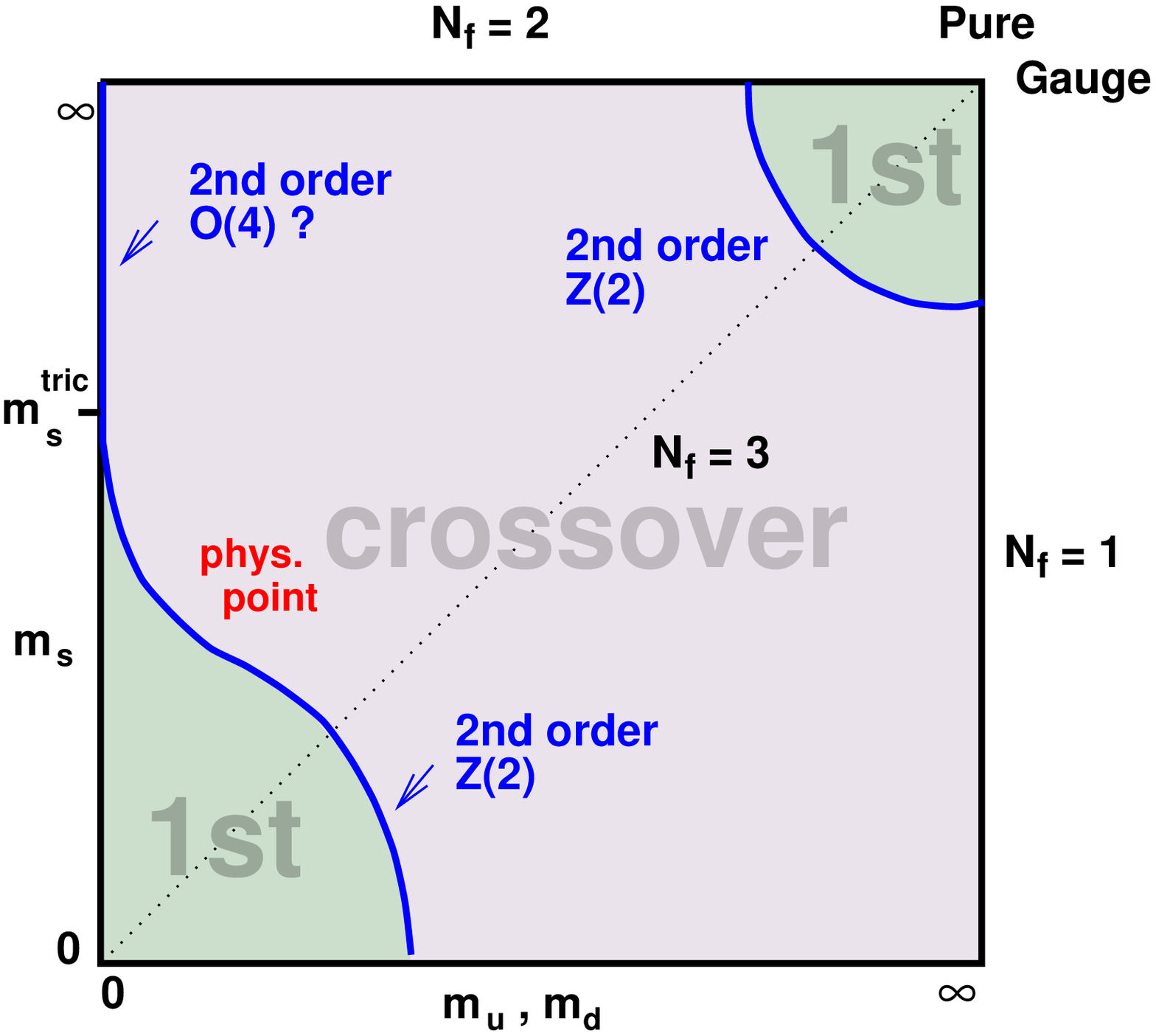}
\hspace*{-0.4cm}
\includegraphics[width=0.40\textwidth]{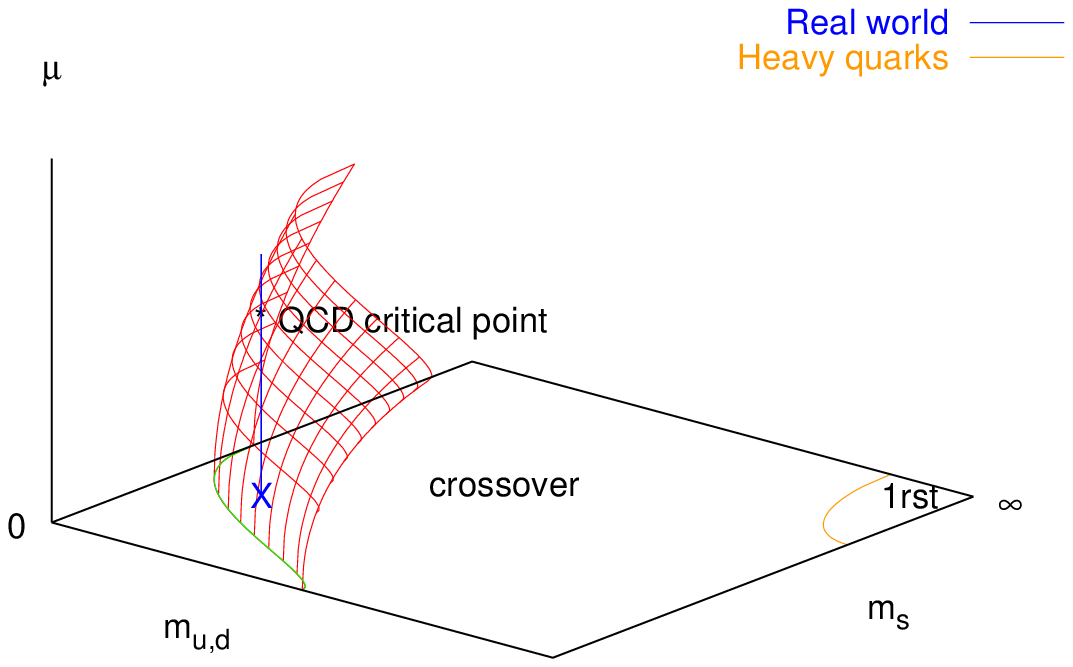}
\put(-122.0,61.0){\tiny $\bullet$}
\hspace*{-0.4cm}
\includegraphics[width=0.40\textwidth]{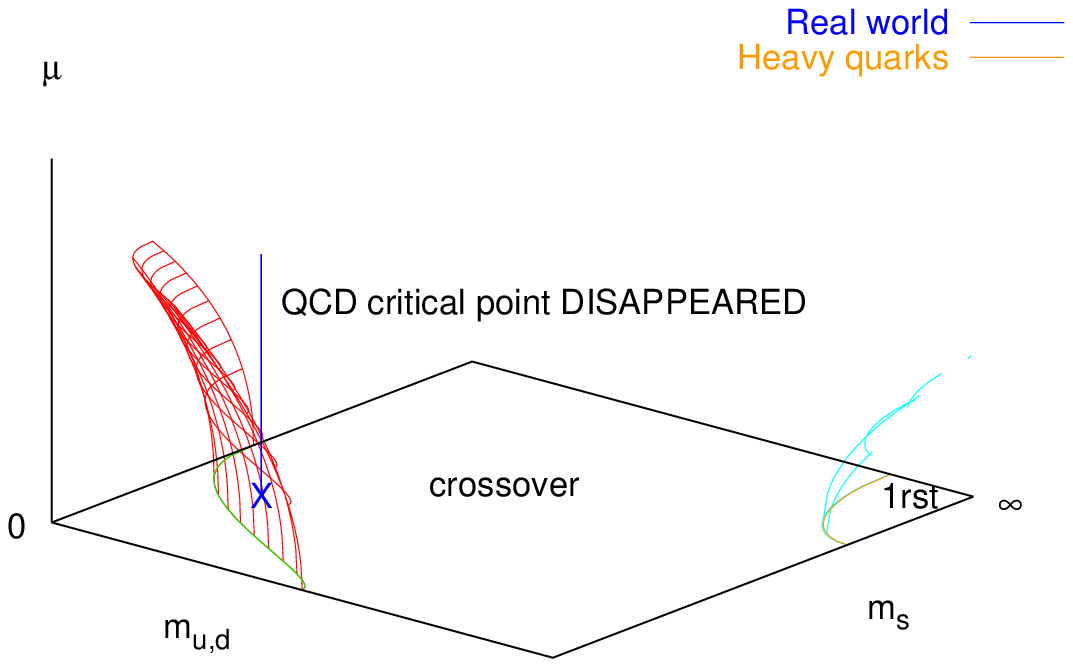}
}
\vspace*{-0.4cm}
{\bf\color{red} \hspace*{4.95cm} $c_1 > 0$ \hspace*{4.5cm} $c_1 < 0$} 
\label{Fig1abc}
\caption{({\em Left}) Schematic phase transition behavior of $N_f=2+1$ QCD
for different choices of quark masses $(m_{u,d},m_s)$ at $\mu=0$.
({\em Middle, Right}) Critical surface swept by the chiral critical line
as $\mu$ is turned on. Depending on the sign of the curvature $c_1$, a QCD chiral
critical point is present or absent~\cite{mu2}. For heavy quarks the curvature
has been determined~\cite{heavy} and the first-order region shrinks with $\mu$.}
\end{center}
\end{figure}

The fundamental importance of the phase diagram of QCD, as a function of temperature $T$
and quark chemical potential $\mu$, makes it the object of several current lattice investigations.
It depends sensitively on the $u,d,s$ quark masses. At $\mu=0$, Fig.~1 ({\em Left}) summarizes
the prevalent understanding of the order of the finite-temperature quark-hadron transition
as a function of $m_u=m_d$ and $m_s$. 
The physical point lies in the crossover region, separated from the chiral, first-order region
by a second-order {\em chiral critical line}. While the $\mu=0$ situation is far from settled, it
can in principle be resolved by manageable increases in computer resources. When $\mu\neq 0$,
the complex nature of the fermion determinant makes the matter much worse. While finite-$\mu$
results, including the location of the QCD critical point, have been obtained by reweighting
$\mu=0$ data~\cite{FK}, assessing the reliability of these results is a challenge in itself~\cite{FKcritique}.
It appears that the only information that can be obtained reliably (i.e. performing thermodynamic
and continuum extrapolations) in principle, barring an algorithmic breakthrough, is the
Taylor expansion of thermodynamic observables in $(\mu/T)$ about $\mu=0$. This makes the
detection of a finite-$\mu$ critical point, characterized by a singularity in the free energy,
particularly difficult. \\
To circumvent this problem, our strategy consists of Taylor-expanding the surface swept by the
chiral critical line of Fig.~1 ({\em Left}). The Taylor expansion of a generic quark mass $m_c$ 
on the {\em chiral critical surface}, and the associated transition temperature $T_c$, can be written as:
\ba
\label{be}
\frac{T_c(m,\mu)}{T_c(m^c_0,0)}&=&1+\sum_{k,l=1} \alpha_{kl}
\left(\frac{m-m^c_0}{\pi T_c}\right)^k
\left(\frac{\mu}{\pi T_c}\right)^{2l},\\
\frac{m_c(\mu)}{m_c(0)}&=&1+\sum_{k=1} c_{k} \left(\frac{\mu}{\pi T_c}\right)^{2k}.
\label{mc}
\ea

The sign of $c_1$ governs the small-$\mu$ behaviour, as illustrated Fig.~1.
Our first results~\cite{mu2}, for the $N_f=3$ $(m_s=m_{u,d})$ theory on an $8^3\times 4$
lattice, favored a negative value for $c_1$.
In \cite{LAT07}, we presented a new numerical method to obtain the $c_k$'s.
Here, we combine the two methods and report on our progress towards determining $c_1$ and higher Taylor coefficients
({\em i}) on larger lattices; ({\em ii}) for the $N_f=2+1$ theory with physical $m_s$;
({\em iii}) for the $N_f=3$ theory on a finer, $N_t=6$, lattice.

\section{Extracting the $\mu$-dependence of the critical point}

On the lattice, the Taylor expansion (\ref{mc}) is replaced by that of dimensionless observables:
\ba
\label{bc}
\beta_c(am,a\mu) & = & \beta_c(am^c_0,0)+ \sum_{k,l=1} c_{kl}\, (am-am^c_0)^k \, (a\mu)^{2l}, \\
\label{mclat}
am^c(a\mu)& = & am^c_0 + \sum_{k=1} c'_k \, (a\mu)^{2k}\;.
\ea
To differentiate between crossover, second- and first-order transitions, we monitor the Binder
cumulant of the quark condensate:
\be
B_4 \equiv \frac{\langle (\delta \bar{\psi}\psi)^4 \rangle}{\langle (\delta \bar{\psi}\psi)^2 \rangle^2},
\quad \delta \bar{\psi}\psi = \bar{\psi}\psi - \langle \bar{\psi}\psi \rangle,
\ee
when $\langle (\delta \bar{\psi}\psi)^3 \rangle = 0$.
On the chiral critical surface, $B_4$ takes value 1.604 as dictated by the $3d$ Ising universality class.
It can be expanded as:
\be
B_4(am,a\mu)=1.604+\sum_{k,l=1}b_{kl}\, (am-am^c_0)^k(a\mu)^{2l}\;,
\label{bseries}
\ee
with coefficients satisfying the scaling behaviour 
$b_{kl}(L)=f_{kl} L^{(k+l)/\nu}$
for large $L$.
Having measured the first few $b_{kl}$'s by the methods of Sec.~3, we can reconstruct the $c'_k$'s 
eq.(\ref{mclat}) as:
\ba 
\label{der1}
c'_1&=&\frac{d\,am^c}{d(a\mu)^2}=-\frac{\partial B_4}{\partial (a\mu)^2}
\left(\frac{\partial B_4}{\partial am}\right)^{-1}=-\frac{b_{01}}{b_{10}} \quad , \\
c'_2&=&
\frac{1}{2!}\,\frac{d^2\,am^c}{d[(a\mu)^2]^2}
=-\frac{1}{b_{10}}(b_{02}+b_{11}c'_1+b_{20}{c'_1}^2) \quad .
\label{der2}
\ea
and finally $c_1$ and $c_2$ as:
\ba
c_1&=&\frac{\pi^2 }{N_t^2}\, \frac{c_1'}{am_0^c}
+\frac{1}{T_c(m^c_0,0)}\frac{dT_c(m^c(\mu),\mu)}{d(\mu/\pi T)^2}, \\
c_2&=&
\frac{\pi^4}{N_t^4}\,\frac{c'_2}{am^c_0}
-\frac{\pi^2}{N_t^2}\,\frac{c'_1}{am^c_0}\,
\frac{1}{T_c(m^c_0,0)}\frac{dT_c(m^c(\mu),\mu)}{d(\mu/\pi T)^2}
+\frac{1}{2T_c(m^c_0,0)}\,\frac{d^2T_c(m^c(\mu),\mu)}{d[(\mu/\pi T)^2]^2}\;.
\label{conv2}
\ea

\section{Two methods to measure $B_4$ derivatives}

\begin{figure}[t]
\begin{center}
\includegraphics[width=0.49\textwidth]{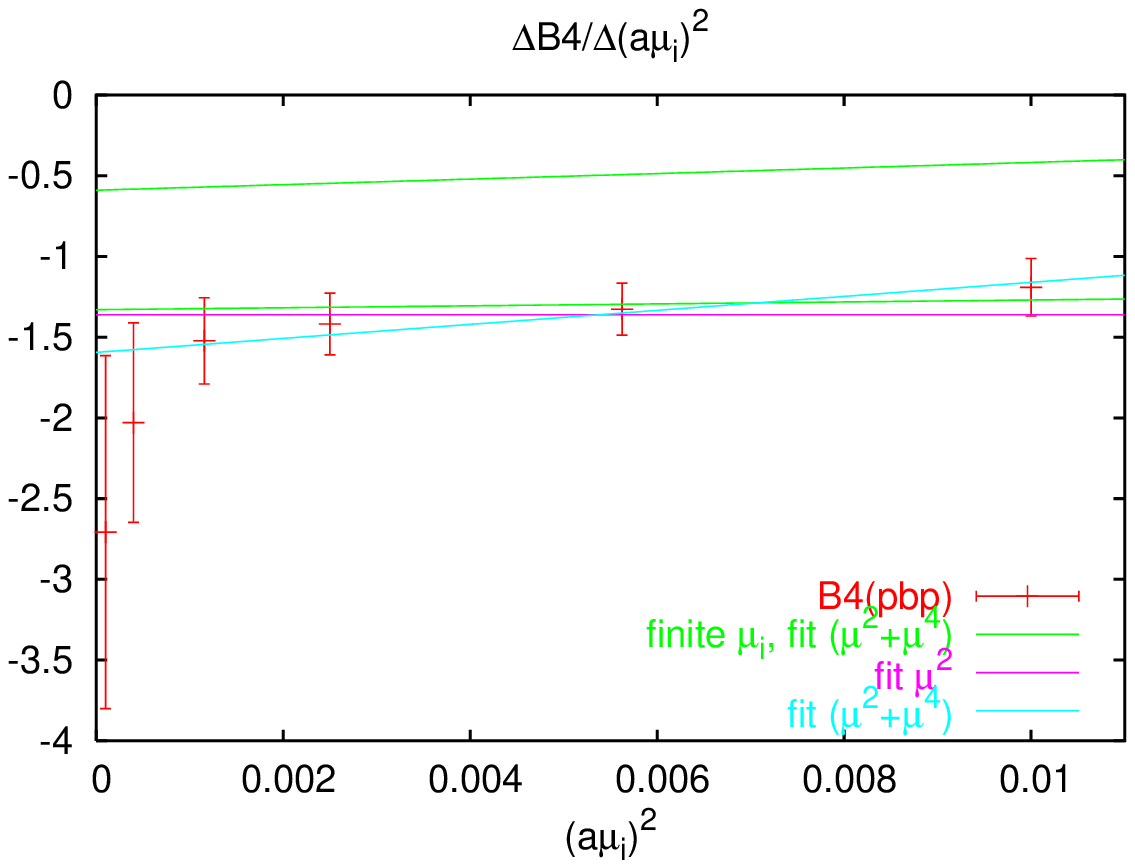}
\includegraphics[width=0.49\textwidth]{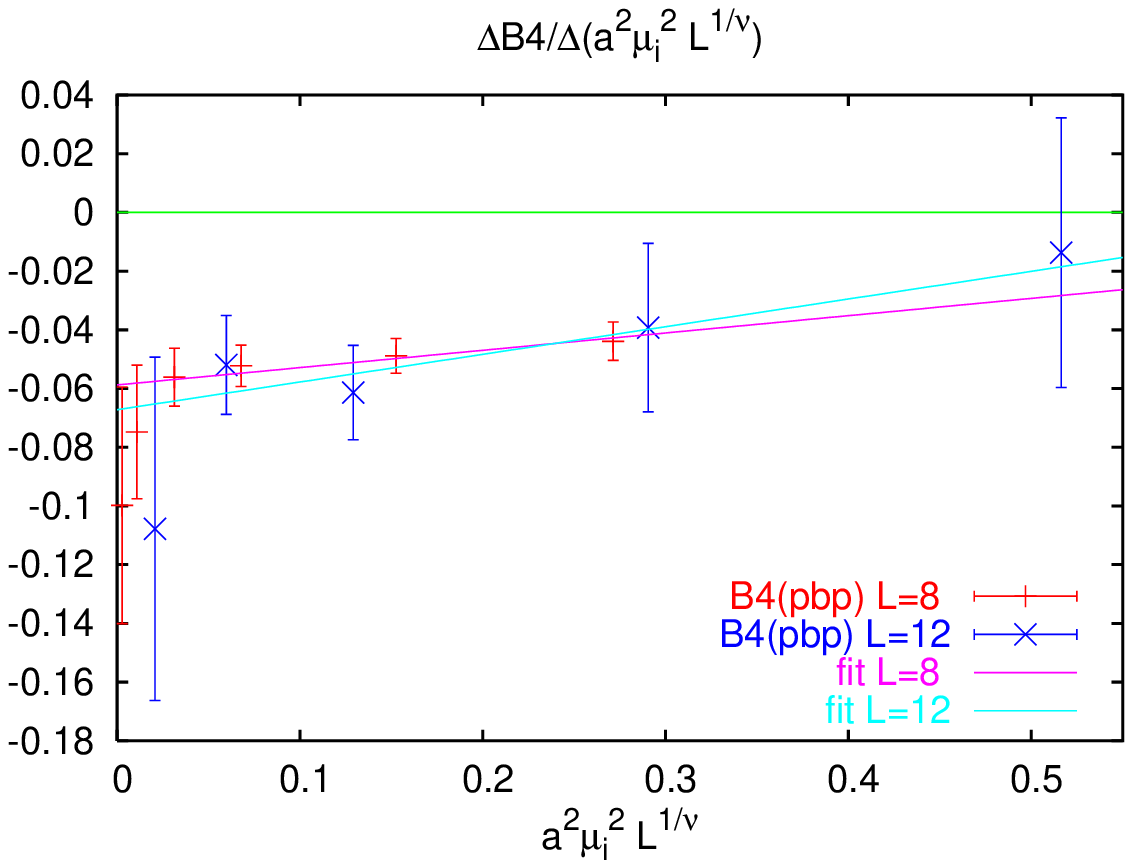}
\label{Fig2ab}
\caption{({\em Left}) Comparison of two methods of measuring $\partial B_4/\partial(a\mu_I)^2$
on an $8^3\times 4$ lattice. The broad error band is the fit to imaginary $\mu$ data; the
data points show the reweighted finite difference quotients, obtained with about 4 times fewer
statistics. ({\em Right}) Finite-size scaling test: data obtained on $8^3$ and $12^3\times 4$
lattices show good consistency with the $3d$ Ising universality class.}
\end{center}
\end{figure}

$B_4$ varies steeply with the quark mass, and $b_{10}, b_{20}$ in eq.(\ref{bseries}) can be
obtained straightforwardly from fits of $B_4$ measured at $\mu=0$ for different quark masses~\cite{mu2}.
Measuring the variation of $B_4$ with $\mu$ is another matter: $B_4$ is a noisy quantity,
its variation is small, and simulating at non-zero (real) $\mu$ is not feasible. We have used two
different, complementary methods to bypass these difficulties~\cite{LAT07}: \\
{\bf 1}. We perform simulations at several imaginary values $\mu=i\mu_i$, where the sign problem is
absent, and fit our measurements of $B_4(\mu_i)$ with a truncated Taylor series in $\mu^2$. \\
{\bf 2}. We perform simulations at $\mu=0$, reweight to small values $\mu=i\mu_i$, and measure the
finite difference quotients $\Delta B_4 / \Delta (a\mu)^2$, with
\be
\lim_{\Delta(a\mu^2)\rightarrow 0}\frac{\Delta B_4}{\Delta (a\mu)^2}=
\left. \frac{\partial B_4}{\partial (a\mu)^2}\right|_{\mu=0} \quad .
\ee
A comparison between the two methods is provided Fig.~2 ({\em Left}), on an $8^3\times 4$ lattice for
$N_f=3$. The error band is the fit to the finite-$\mu_i$ data (method {\bf 1}). The data points are 
the finite-difference quotients (method {\bf 2}). Consistency between the two methods is observed.
The second method is clearly more efficient, since the statistics is only 1/4 of the other.
This efficiency can be traced to the strong cancellation of statistical fluctuations when measuring
$\Delta B_4$ on the $\mu=0$ and the reweighted ensemble. Reweighting itself is done stochastically
with a Gaussian-distributed vector $\eta$, since the reweighting factor is
\be
\rho(\mu_1,\mu_2)=\frac{\det^{N_f/4} \Dslash(U,\mu_2)}
{\det^{N_f/4}\Dslash(U,\mu_1)} 
= \left\langle \exp\left(-|\Dslash^{-N_f/8}(\mu_2) 
\Dslash^{+N_f/8}(\mu_1) \eta|^2 + |\eta|^2 \right) \right\rangle_\eta .
\ee
Note the small values of $(a\mu_i)^2$ in Fig.~2 ({\em Left}): they guarantee a good overlap between
the $\mu=0$ Monte Carlo ensemble and the reweighted $\mu=i\mu_i$ ensemble, and small fluctuations in $\rho$.
 
\begin{figure}[t]
\begin{center}
\includegraphics[width=0.75\textwidth]{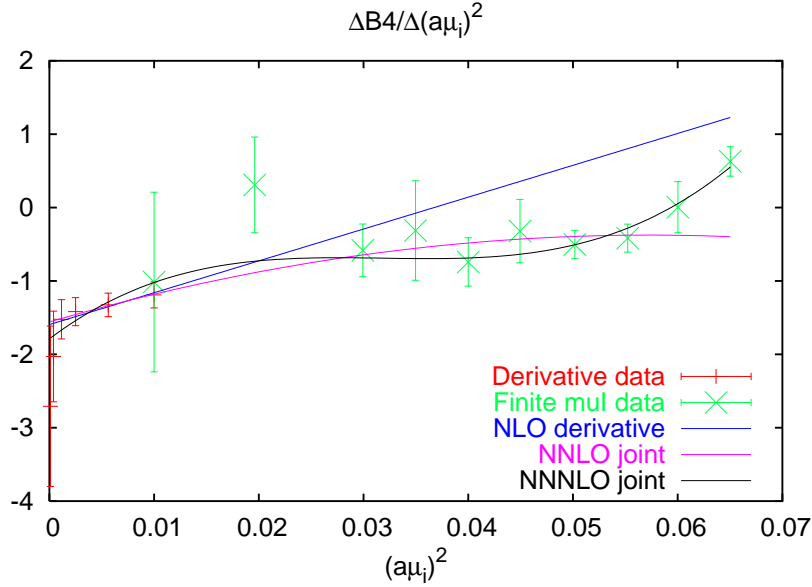}
\label{Fig3}
\caption{Combining the two methods: the $(a\mu_i)^2\leq 0.01$ data come from $\mu=0$ reweighting, the
$(a\mu_i)^2\geq 0.01$ from direct $\mu_i\neq 0$ simulations, all at $am=0.0265$. Data at larger $\mu_i$ clearly
fall below the ${\cal O}(\mu_i^4)$ contribution, indicating a negative $\mu_i^6$-term. The quality
of the cubic, S-shape fit favors a positive $\mu_i^8$-term. After rotation to real $\mu$, all terms 
contribute to increasing $B_4$, i.e. pushing the system in the crossover region.
}
\end{center}
\end{figure}

Since our $8^3$ lattice is not very large ($m_\pi L \sim 3.4$), we performed a finite-size scaling check
by comparing with a $12^3\times 4$ lattice. Fig.~2 ({\em Right}) shows nice consistency with the expected
large volume universal behaviour, not only for the $y$-axis intercept yielding $b_{01}$, but also for
the slope yielding $b_{02}$. The result ($b_{02} > 0$ like $b_{01}$) reinforces the finding that the transition
weakens and turns into a crossover
(i.e. $B_4$ increases) as $\mu$ is turned on (see eq.~(\ref{bseries})).

Finally, we can combine the data from our two methods, since the simulations were performed independently
and cover different ranges of $\mu_i$. A combined fit of the $am = 0.0265$ data Fig.~3 shows that 
$(B_4(a\mu_i) - B_4(\mu=0))/(a\mu_i)^2$ 
is an alternating series in $(a\mu_i)^2$~\cite{bari}. The fit gives
\be
B_4(a\mu_i) = B_4(\mu=0) - 1.79(14) (a\mu_i)^2 + 108(27) (a\mu_i)^4 - 3438(933) (a\mu_i)^6 + 35954(8876) (a\mu_i)^8
\ee
with a $\chi^2$/d.o.f. of 0.57. The large values of higher-order coefficients indicate that higher-order terms become important
when $\mu/T \gsim 0.5$. However, after rotation to real $\mu$, they all tend to {\em increase} $B_4$, pushing the
system deeper in the crossover region.
This only increases the validity of the exotic scenario Fig.~1 ({\em Right})
up to larger values of $\mu/T$. Conservatively, we trust only the ${\cal O}(\mu^2)$ and ${\cal O}(\mu^4)$ terms.
After continuum conversion following eqs.(\ref{der1}-\ref{conv2}), 
our final result for $N_f=3$ on coarse, $N_t=4$, lattices reads~\cite{mu4}:
\be
\frac{m_c(\mu)}{m_c(0)}=1-3.3(3)\left(\frac{\mu}{\pi T}\right)^2-47(20)\left(\frac{\mu}{\pi T}\right)^4
-\ldots
\label{resu}
\ee

\section{Towards the $N_f=2+1$ continuum limit}

We are currently investigating two reasons why our result eq.(\ref{resu}) could change
qualitatively as we consider real QCD. The sign of the curvature could change as we move along
the critical line away from the degenerate $N_f=3$ case. It could also change as we take the 
continuum limit.

The first possibility appears unlikely given our current results Fig.~4 ({\em Left}), where 
$m_s$ is given its physical value on the $N_t=4$ critical line determined in \cite{mu2}
(see Fig.~4 ({\em Right})). 
Since our pions are lighter than in nature, large lattices are required and thereby large computer
resources. This is achieved, like for the $N_f=3, 8^3\times 4$, method {\bf 2} case above, by dispatching
our simulations over the computing Grid. Many independent Monte Carlo runs are performed, all at $\mu=0$,
over a range of temperatures near $T_c$, using prioritized scheduling. Current statistics reach
600k thermalized configurations.

\begin{figure}[t]
\begin{center}
\includegraphics[width=0.512\textwidth]{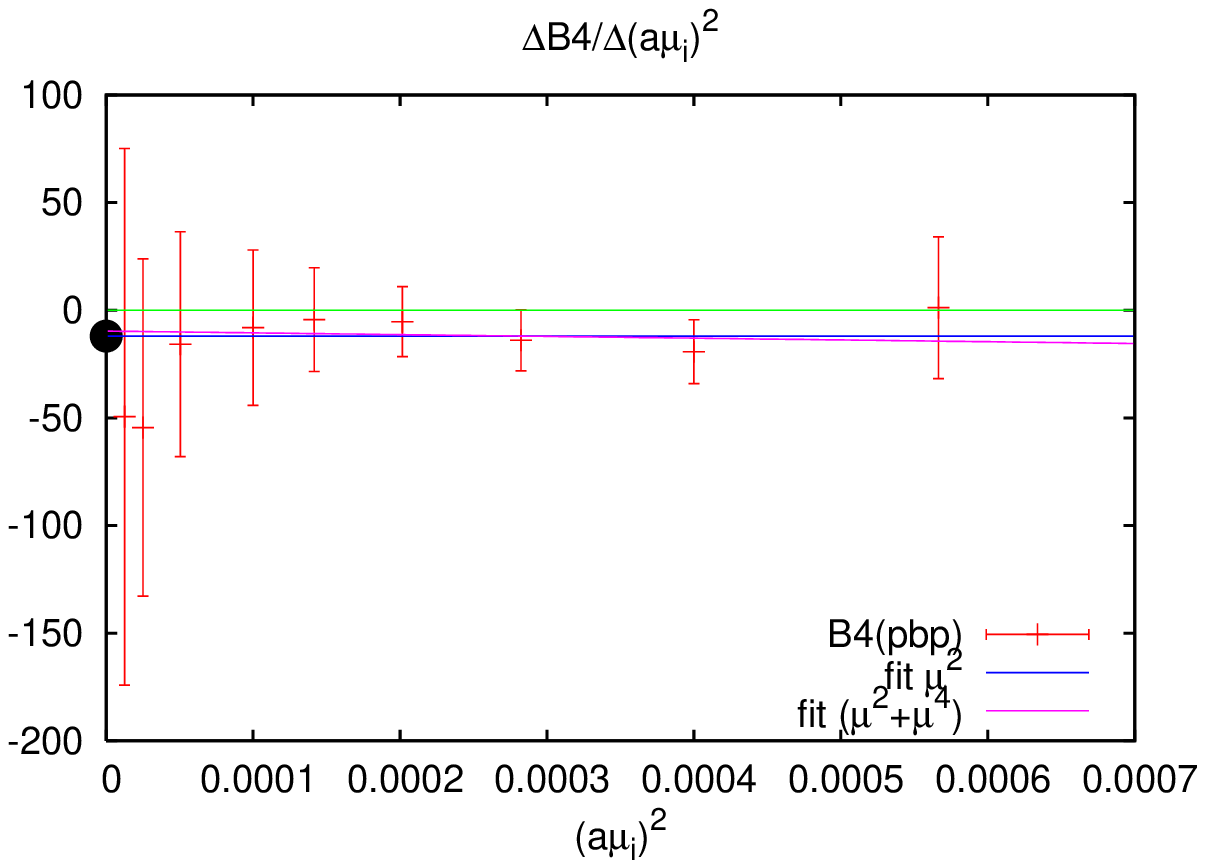}
\includegraphics[width=0.475\textwidth]{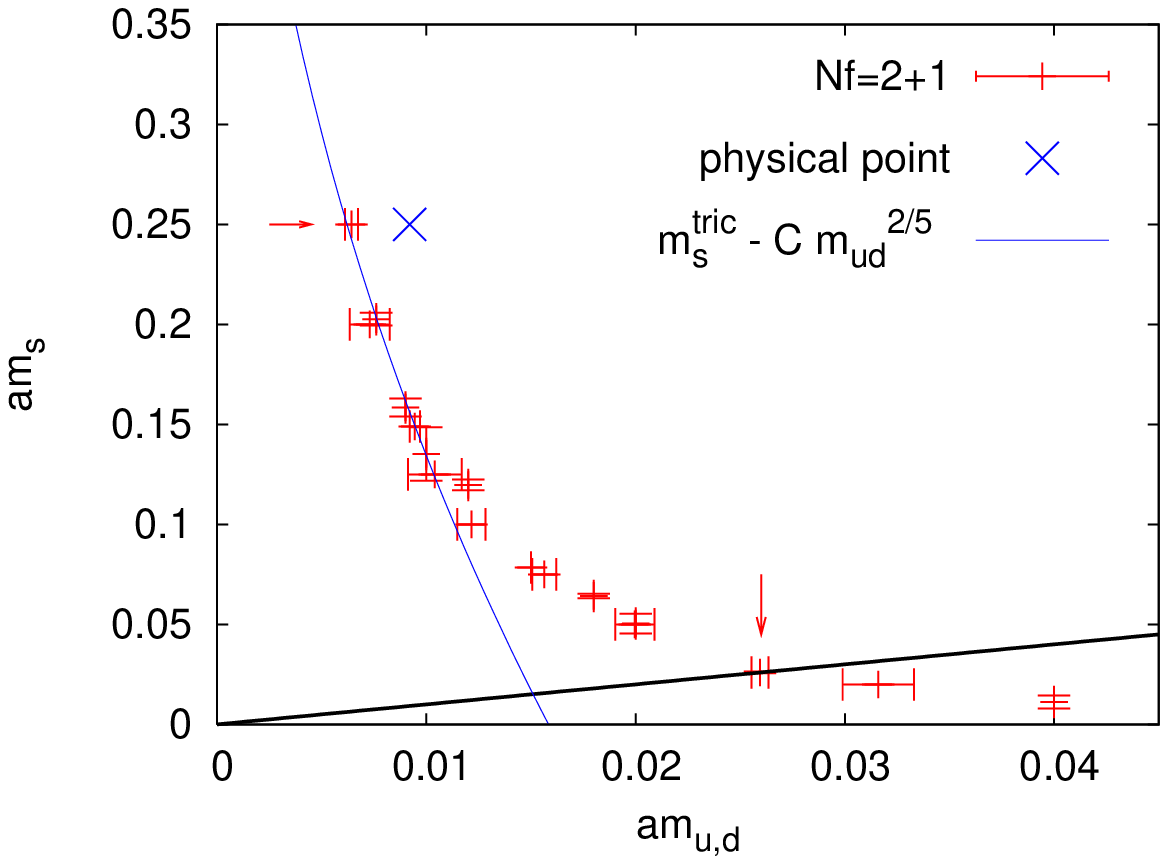}
\label{Fig4ab}
\caption{Work in progress: ({\em Left}) $N_f=2+1$ on a $16^3\times 4$ lattice.
The simulation point, indicated by the leftmost arrow ({\em Right}), lies to the
left of the physical point, implying that our pions are lighter than in nature. 
}
\end{center}
\end{figure}

The effect of a finer lattice is studied by simulating $18^3\times 6$
lattices with $N_f=3$ degenerate flavors. The current results, Fig.~5 ({\em Left}),
give opposite signs for $b_{01}$ using a leading or subleading order fit. While the
sign of the curvature $c_1$ is consequently not clear, one can already say that 
$|c_1|$ is not large, ${\cal O}(20)$ or less. Thus, the critical surface is almost vertical.

In addition, another qualitative effect takes place: the $\mu=0$ critical line, and
thereby the whole chiral critical surface, moves towards the origin as $a \to 0$.
For instance, the $N_f=3$ pion mass on the critical line drops from $1.680(4) T_c$
to $0.954(12) T_c$ going from $N_t=4$ to $N_t=6$ lattices~\cite{LAT07}.
The first-order region, in physical units, shrinks dramatically as $a \to 0$.
To compensate this effect and maintain a critical point for real QCD at small chemical
potentials $\mu/T \lesssim 1$, a large positive curvature $c_1$ would be needed.
We presently do not see it.

Finally, we note that effective models like PNJL~\cite{Fuku} or linear sigma model~\cite{Kapusta},
with simple modifications,
can reproduce the qualitative features of the chiral critical surface which we observe. 
Nevertheless, let us stress again that our study concerns only the {\em chiral}
critical surface, swept by the $\mu=0$ {\em chiral} critical line as the chemical potential
is turned on. Our results do not preclude other phase transitions, not connected to the
chiral one.

\begin{figure}[t]
\centerline{
\includegraphics[width=0.40\textwidth]{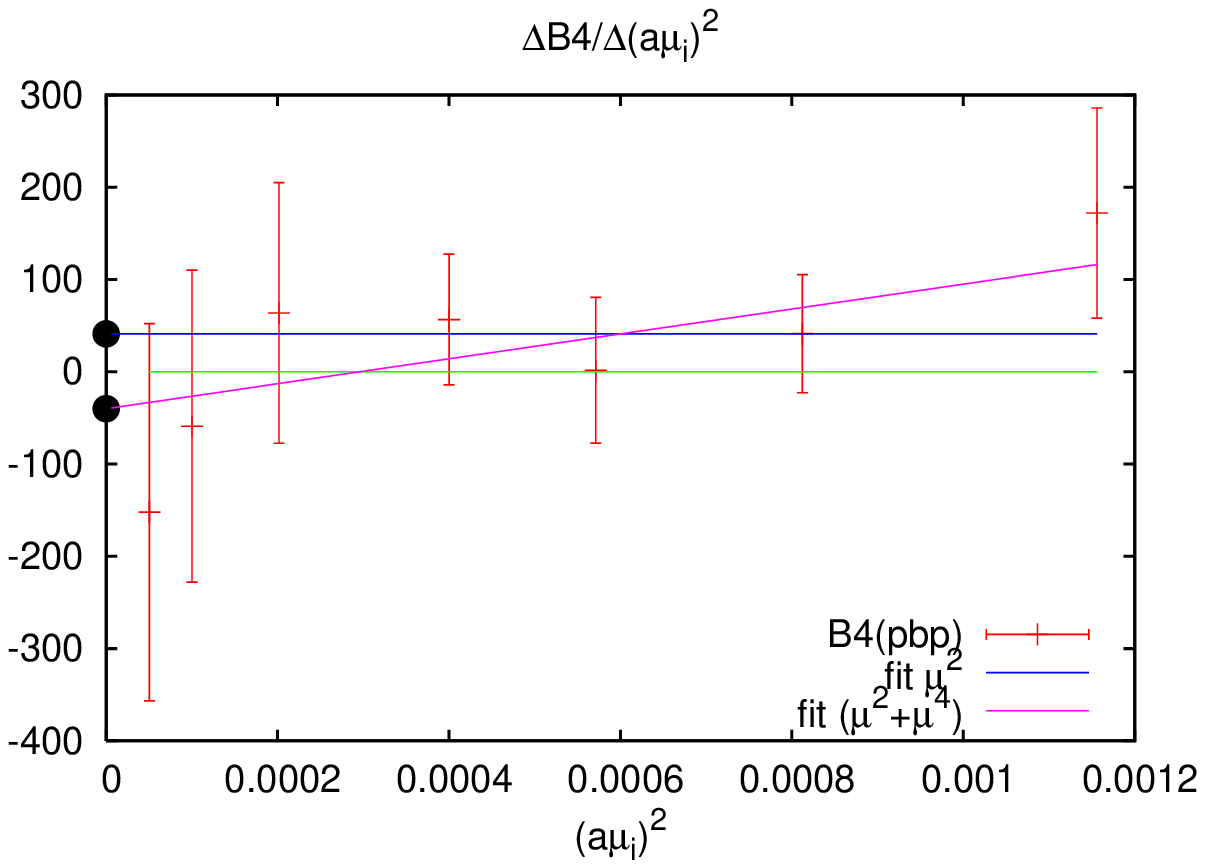}
\hspace*{-0.5cm}
\includegraphics[width=0.38\textwidth]{3dphasediag_3.eps}
\put(-116.0,56.0){\tiny $\bullet$}
\put(-124,56){\color{blue}\bf $\uparrow$}
\put(-135,31){\color{blue}\bf $\longleftarrow$}
\hspace*{-0.7cm}
\includegraphics[width=0.38\textwidth]{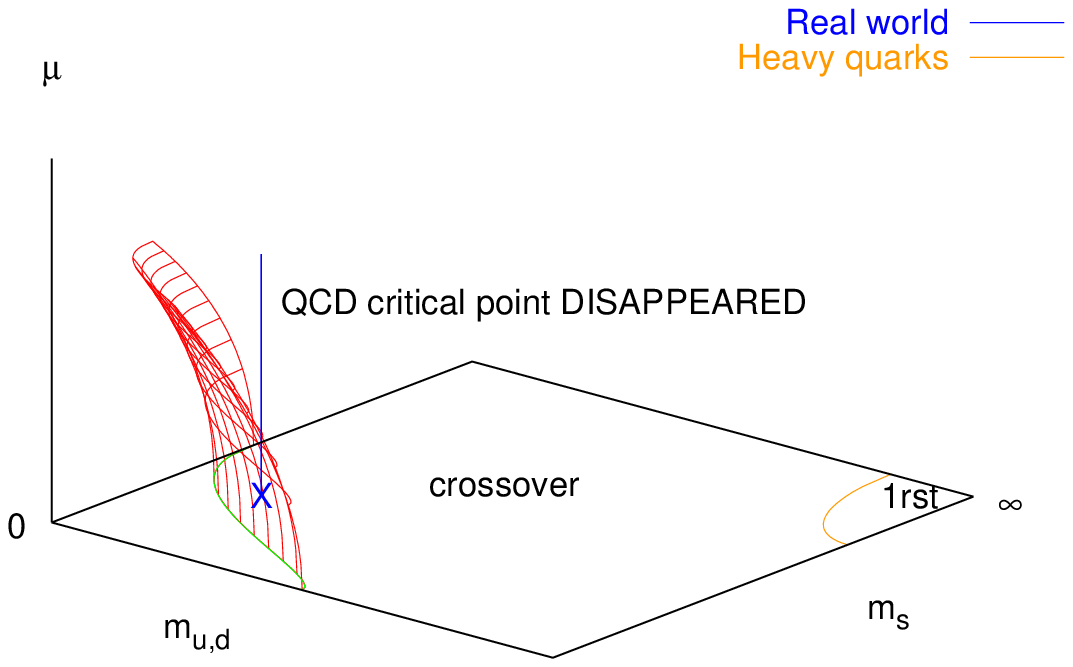}
\put(-135,31){\color{blue}\bf $\longleftarrow$}
}
\label{Fig5abc}
\begin{center}
\caption{Work in progress:({\em Left}) $N_f=3$ on an $18^3\times 6$ lattice.
The sign of the intercept $b_{01}$ depends on the fitting ansatz.
({\em Middle}) As the lattice spacing is reduced, the critical surface moves towards the origin.
If a critical point exists, its location moves to larger $\mu$.
({\em Right}) If the curvature $c_1$ is negative, higher-order terms must increase
in magnitude for a critical point to occur.
}
\end{center}
\end{figure}

\section*{Acknowledgements:}
This work is partially supported by the German BMBF, project
{\em Hot Nuclear Matter from Heavy Ion Collisions
and its Understanding from QCD}, No.~06MS254.
We thank the Minnesota Supercomputer Institute 
for providing computer resources,
and the CERN IT/GS group for their invaluable assistance and collaboration 
using the EGEE Grid for part of this project.
We acknowledge the usage of EGEE resources (EU project under contracts
EU031688 and EU222667). Computing resources have been contributed by a number
of collaborating computer centers, most notably HLRS Stuttgart (GER), NIKHEF (NL), CYFRONET (PL),
CSCS (CH) and CERN.

\end{document}